\documentclass[onecolumn,amsmath,amssymb,12pt,superscriptaddress,nofootinbib]{revtex4}

\def\gsim{ \lower .75ex \hbox{$\sim$} \llap{\raise .27ex \hbox{$>$}} }

\def\lsim{ \lower .75ex \hbox{$\sim$} \llap{\raise .27ex \hbox{$<$}} }

\usepackage{graphicx}
\usepackage{dcolumn}
\usepackage{bm}
\input epsf
\def\gsim{ \lower .75ex \hbox{$\sim$} \llap{\raise .27ex \hbox{$>$}} }
\def\lsim{ \lower .75ex \hbox{$\sim$} \llap{\raise .27ex \hbox{$<$}} }

\usepackage{graphicx}
\usepackage{dcolumn}
\usepackage{bm}

\newcommand{\nn}{\nonumber}
\newcommand{\be}{\begin{equation}}
\newcommand{\ee}{\end{equation}}
\newcommand{\bea}{\begin{eqnarray}}
\newcommand{\eea}{\end{eqnarray}}
\renewcommand{\d}{\mathrm{d}}
\def\a{\alpha}

\def\d{\delta}

\def\p{\phi}

\def\th{\theta}
\def\tb{\bar{\theta}}

\def\pt{\partial}

\def\Pd{\Phi^\dag}

\usepackage{tikz}
\usetikzlibrary{arrows,decorations.markings,mindmap,patterns}

\newlength{\oldoddsidemargin}
\oldoddsidemargin=\oddsidemargin \oddsidemargin=\evensidemargin
\evensidemargin=\oldoddsidemargin

\makeatletter
\def\cleardoublepage{\clearpage\if@twoside \ifodd\c@page\else
   \hbox{}
   \thispagestyle{empty}
   \newpage
   \if@twocolumn\hbox{}\newpage\fi\fi\fi}
\makeatother \clearpage{\pagestyle{plain}\cleardoublepage} 
\usepackage{tabularx}
\usepackage{longtable}
\usepackage{multirow}
\usepackage{graphicx, color}
\usepackage{subfigure}
\usepackage{amsmath}
\usepackage{amsfonts}
\usepackage{amssymb}
\usepackage[english,german]{babel}
\usepackage{bbding}

\usepackage{float}
\usepackage{dsfont}   
\usepackage{hyperref}
\usepackage{mathrsfs} 
\usepackage{makeidx}
\usepackage{listings}
\makeindex
\usepackage{setspace}
\newcolumntype{C}[1]{>{\centering\arraybackslash}b{#1}}
\newcolumntype{D}[1]{>{\centering\arraybackslash}b{#1}}
\newcolumntype{R}[1]{>{\raggedleft\arraybackslash}b{#1}}                                                                                                                                                                                                                                                                                                                                                                                                                                                                                                                                                                                                                                                                                                                                                                                                                                                                                                                                                                                                                                                                                                                                                                                                                                                                                                                                                                                                                                                                                                                                                                                                                                                                                                                                                                                                                                                                                                                                                                                                                                                                                                                                                                                                                                                                                                                                                                                                                                                                                                                                                                                                                                                                                                                                                                                                                                                                                                                                                                                                                                                                                                                                                                                                                                                                                                                                                                                                                                                                                                                                                                                    
\newcolumntype{M}[1]{>{\centering\arraybackslash}m{#1}}

\def\p{\partial}
\def\d{\mathrm d}

\def\bea{\begin{eqnarray}}
\def\eea{\end{eqnarray}}

\def\I{\mathrm{i}}

\def\bsm{\left( \!\begin{smallmatrix}}
\def\esm{\end{smallmatrix} \!\right)}

\renewcommand{\nn}{\nonumber}

\numberwithin{equation}{section}

\renewcommand{\a}{\alpha}

\renewcommand{\th}{\theta}

\newcommand{\ba}{\begin{align}}
\newcommand{\ea}{\end{align}}

\begin{document}
\selectlanguage{english}
\allowdisplaybreaks
\begin{titlepage}

\title{Supersymmetric Cubic Galileons Have Ghosts}

\author{Michael Koehn}
\email[]{michael.koehn@aei.mpg.de}
\author{Jean-Luc Lehners}
\email[]{jlehners@aei.mpg.de}

\affiliation{Max--Planck--Institute for Gravitational Physics (Albert--Einstein--Institute), 14476 Potsdam, Germany}
\author{Burt Ovrut}
\email[]{ovrut@elcapitan.hep.upenn.edu}
\affiliation{Department of Physics, University of Pennsylvania,\\ 209 South 33rd Street, Philadelphia, PA 19104-6395, U.S.A.}

\begin{abstract} 

\vspace{.3in}
\noindent Galileons are higher-derivative theories of a real scalar which nevertheless admit second order equations of motion. They have interesting applications as dark energy models and in early universe cosmology, and have been conjectured to arise as descriptions of brane dynamics in string theory. In the present paper, we study the bosonic sector of globally $N=1$ supersymmetric extensions of the cubic Galileon Lagrangian in detail. Supersymmetry requires that the Galileon scalar now becomes paired with a second real scalar field. We prove that the presence of this second scalar causes the equations of motion to become higher than second order, thus leading to the appearance of ghosts. 
We also analyze the energy scales up to which, in an effective field theory description, the ghosts can be tamed.

\end{abstract}
\maketitle

\thispagestyle{empty}

\end{titlepage}

\section{Introduction}

Galileon theories of a real scalar field are special because they have two-derivative equations of motion despite having higher-derivative Lagrangians. They are a sub-class of the most general scalar theories with two-derivative equations of motion, known as Horndeski's theories \cite{Horndeski:1974wa} (see also \cite{Deffayet:2011gz}). The ``standard'' Galileons \cite{Nicolis:2008in} have the additional property that in the equations of motion there are precisely two derivatives acting on each field. An immediate consequence is that the standard Galileons are invariant under a so-called Galilean shift symmetry $\phi \rightarrow \phi + c + b_\mu x^\mu$ with $c,b_\mu$ being constants, whence they derive their name. 
Many variants of the original model have been constructed, such as conformal Galileons \cite{Nicolis:2009qm}, DBI Galileons \cite{deRham:2010eu}, Galileons with an internal symmetry \cite{Hinterbichler:2010xn,Goon:2011qf}, bi-Galileons \cite{Deffayet:2010zh,Padilla:2010de} and so on.
The crucial property of all of these theories is that they have equations of motion with no more than two derivatives acting on a field. This helps to evade Ostrogradsky's theorem \cite{Ostrogradsky} -- that is, despite the higher-derivative nature of the Lagrangians, for suitable coefficients of the Galileon Lagrangians these theories {\it do not contain ghosts}.

Galileons have attracted considerable interest due to their rather remarkable properties. For example, they admit de-Sitter-like solutions in the absence of a cosmological constant \cite{Silva:2009km,DeFelice:2010pv,Chow:2009fm} and they lead to a Vainshtein-type screening mechanism so that they can be in agreement with solar system ``fifth force'' constraints while contributing a fifth force on large scales \cite{Vainshtein:1972sx,Deffayet:2001uk}. Moreover, they allow for solutions that violate the null energy condition without leading to the appearance of ghosts \cite{Nicolis:2009qm,Hinterbichler:2012yn}. This last property means that Galileons also have applications to early universe cosmology, allowing the construction of emergent cosmologies (see, for example, the model of Galilean genesis \cite{Creminelli:2010ba}) and non-singular bouncing cosmologies such as new ekpyrotic theory \cite{Buchbinder:2007ad,Buchbinder:2007tw,Buchbinder:2007at,Lehners:2009ja,Lehners:2010fy,Lehners:2011kr} or the matter bounce model \cite{Cai:2012va}. Such alternative models to inflation even play a significant role in eternal inflation \cite{Lehners:2012wz,Johnson:2011aa,Liu:2013xt}.

There exists a suggestive construction of Galileon Lagrangians as the theories describing the dynamics of co-dimension one branes \cite{deRham:2010eu}. This has led people to speculate that Galileons might arise naturally out of string theory and, hence, enjoy a more fundamental status than other higher-derivative terms, in analogy to the Dirac-Born-Infeld action. Brane backgrounds in string theory typically preserve some amount of unbroken supersymmetry. Therefore, if Galileons are to arise from string theory it will be in a supersymmetric context. Hence, it is of importance to study the supersymmetric extensions of Galileon theories. 
In previous work \cite{Khoury:2011da}, it was shown that {\it conformal} Galileons can be made globally $N=1$ supersymmetric--these theories arising naturally as a way of obtaining correct sign spatial gradients in supersymmetric ghost-condensates (see also \cite{Khoury:2010gb,Koehn:2012te}). It was found that the new fields required by supersymmetry (a second real scalar, a spin $\frac12$ fermion and a complex auxiliary field) admit stable, positive-energy fluctuations around specific backgrounds, namely those where the second scalar field is constant. However, possible ghost instabilities associated with vacua with a {\it spacetime dependent} second scalar were not explored.
We will do this in the present paper, restricting our discussion for the most part to the cubic Galileons within the context of four-dimensional global $N=1$ supersymmetry.

To begin, we present {\it complex} scalar Galileons which, when the second scalar is set to zero, reduce to real Galileons of the $L_{3}$, $L_{4}$ and $L_{5}$ type. These possess manifestly two-derivative equations of motion and a Galilean symmetry for the two constituent real scalars fields.
We then show, however, that such complex Galileon theories {\it cannot} be obtained in $N=1$ supersymmetry. 
We next consider the cubic-in-the-field, four-derivative $L_{3}$ Lagrangian, and show that there is a unique possible $N=1$ supersymmetric generalization. However, it is demonstrated that 
this Lagrangian leads to {\it higher-derivative equations of motion}!
An immediate consequence is that, around general backgrounds, this theory admits a ghost, whose existence we explicitly demonstrate.  In the effective field theory context, we then calculate the mass of the ghost and argue that for a sufficiently low cut-off scale the ghost degree of freedom can be safely ignored. In the final technical section before the discussion, we extend our analysis to supersymmetrize the quartic-in-the-field, six-derivative $L_4$ Lagrangian. Here, we do not provide an exhaustive treatment of all possible supersymmetric extensions of $L_4.$ We merely present one possible extension of $L_4,$ which we again find to lead to higher-derivative equations of motion. This time we perform the stability analysis using the canonical Hamiltonian formalism, and explicitly demonstrate the existence of ghosts as well as the unboundedness of the Hamiltonian. Our example illustrates that the appearance of ghosts is rather generic for supersymmetric extensions of the Galileon Lagrangians. However, a recent paper by Farakos {\it et al.} has explicitly demonstrated that it is possible to construct a ghost-free supersymmetric extension of $L_4$ \cite{Farakos:2013fne}. Thus, our no-go result for cubic Galileons does not extend to the quartic Galileons in general, while the status of the quintic Galileons is currently still an open problem. It is notable however that, for once, the inclusion of supersymmetry does not necessarily improve the stability properties of a theory -- quite to the contrary!

We note that we have performed our analysis within the context of global rather than local supersymmetry. However, the generic supersymmetric structure of the higher-derivative scalar field Lagrangians is not substantially altered in the presence of gravity (see {\it e.g.} \cite{Koehn:2012ar,Koehn:2012np}). That is, the existence of ghosts in the $L_{3}$ Galileons will persist when these are coupled to $N=1$ supergravity. Finally, we would like to stress that our results are derived for the ``standard'' Galileon theories. Since the cubic conformal Galileon contains precisely the same cubic term (and in addition a quartic $(\pt\phi)^4$ term) \cite{Nicolis:2009qm}, our results immediately extend to this Lagrangian also. 


\section{Galileons and Complex Fields} \label{ComplexGalileons}

In this and the following two sections, we will focus on the simplest non-trivial Galileon Lagrangian given by \cite{Nicolis:2008in}
\be
L_3 = -\frac{1}{2} (\p\phi)^2 \Box \phi. \label{L3}
\ee
By varying with respect to $\phi,$ one can immediately see that the equation of motion is second order and given by
\be
(\Box \phi)^2 - \phi^{,\mu\nu} \phi_{,\mu\nu} = 0. \label{L3eom}
\ee
Thus, despite the higher-derivative nature of the Lagrangian, the equation of motion is well-behaved and the Cauchy problem is well-posed. In four dimensions, there are two more such Galileon Lagrangians,
\bea
L_4 &=& -\frac12 (\p\phi)^2 \big((\Box \phi)^2 - \phi^{,\mu\nu} \phi_{,\mu\nu} \big), \label{L4} \\ L_5 &=& -\frac12 (\p\phi)^2 \big((\Box\phi)^3 - 3 \Box \phi \phi^{,\mu\nu} \phi_{,\mu\nu} + 2 \phi^{,\mu\nu} \phi_{,\mu\rho} \phi_{,\nu}{}^\rho\big) \label{L5}
\eea
which also lead to second-order equations of motion. For example, the equation of motion for $L_4$ is given by
\be
(\Box\phi)^3 - 3 \Box \phi \phi^{,\mu\nu} \phi_{,\mu\nu} + 2 \phi^{,\mu\nu} \phi_{,\mu\rho} \phi_{,\nu}{}^\rho = 0.
\ee
A detailed discussion of the $L_{4}$ Lagrangian and one of its supersymmetric extensions will be discussed below in the final technical section.

In $N=1$ supersymmetry, scalar field theories can be constructed using chiral superfields $\Phi.$ The lowest component of such a superfield is a {\it complex} scalar $A,$ which can be decomposed into two real scalars as
\be
A = \frac{1}{\sqrt{2}} (\phi + \I \xi). \label{A}
\ee
One consequence is that supersymmetric scalar field actions can always be written as hermitian combinations of $A$ and its complex conjugate $A^*$. 
Motivated by this, {\it but before imposing any supersymmetry condition}, it is of interest to consider the possible extensions of the Galileon Lagrangian \eqref{L3} from the real scalar field $\phi$ to the complex scalar $A$ in \eqref{A}.
Specifically, we are interested in Lagrangians which, when the second real scalar $\xi$ is set to zero, reduce to the Galileon Lagrangian $L_3$ presented in \eqref{L3}. There are, in principle, a large number of such Lagrangians. Here, we do not try to give an exhaustive treatment--since, as we show in the next section, most will be incompatible with supersymmetry. Rather, we will illustrate using two concrete examples that, even though by construction these extended Lagrangians contain the $L_3$ Lagrangian for $\phi,$ the properties of the second scalar $\xi$ can vary considerably, and it is in no way guaranteed that the second scalar also shares the desired Galilean symmetries. Having established this,  we will then--in Section III--move on to supersymmetry (where we will give a completely exhaustive treatment) in order to determine which such complex scalar field generalizations of $L_3$ supersymmetry allows.

Our first example of a generalization of \eqref{L3} from the real scalar $\phi$ to a complex scalar field $A$ is straightforward. It is obtained simply by replacing $\phi \rightarrow \sqrt{2} A$ and then taking the real part. For $L_3$ above, this amounts to considering the Lagrangian
\be
L_3^{\mathbb{C}} = -\frac{1}{\sqrt{2}} (\p A)^2 \Box A + h.c. \ ,
\label{hello1}
\ee
where $h.c.$ stands for ``hermitian conjugate''. It is then evident that the resulting equations of motion are still second order, since they are given by
\be
(\Box A)^2 - A^{,\mu\nu} A_{,\mu\nu} = 0, \qquad (\Box A^*)^2 - A^{*,\mu\nu} A^*_{,\mu\nu} = 0.
\ee
In terms of the real scalars $\phi$ and $\xi,$ the Lagrangian and equations of motion are
\bea
L_3^{\mathbb{C}} &=& -\frac12 \big((\p\phi)^2 \Box \phi - (\p\xi)^2 \Box \phi -2 \p\phi \cdot \p\xi \Box \xi\big), \label{L3C} \\
0 &=& (\Box \phi)^2 - \phi^{,\mu\nu} \phi_{,\mu\nu} - (\Box \xi)^2 + \xi^{,\mu\nu} \xi_{,\mu\nu}, \\
0 &=& \Box\phi \Box\xi - \phi^{,\mu\nu} \xi_{,\mu\nu},
\eea
clearly exhibiting that we now have a coupled two-field Galileon system. Not only are the equations of motion of second order, but both fields admit independent Galileon-type shift symmetries $\phi \rightarrow \phi + c^{(\phi)} + b^{(\phi)}_{\mu} x^\mu$ and $\xi \rightarrow \xi + c^{(\xi)} + b^{(\xi)}_{\mu} x^\mu$ respectively.

However, using a second concrete example, we now demonstrate that other extensions of the $L_3$ Lagrangian to complex scalar field $A$ do {\it not} necessarily lead to second-order equations of motion. To illustrate this important point, consider the action
\bea
\tilde{L}_{3}^\mathbb{C} &=& -\frac{1}{\sqrt{2}} \p A \cdot \p A^* \Box A + h.c. \label{again}\\
&=& -\frac12 \big((\p\phi)^2 \Box \phi + (\p\xi)^2 \Box \phi\big),
\eea
leading to the equations of motion
\bea
0 &=& (\Box \phi)^2 - \phi^{,\mu\nu} \phi_{,\mu\nu} -\xi^{,\mu\nu} \xi_{,\mu\nu} - \xi_{,\mu} \xi_{,\nu}{}^{\nu\mu},\\
0 &=& \Box \xi \Box \phi + \xi_{,\mu} \phi_{,\nu}{}^{\nu\mu}.
\eea
Clearly, these are higher-order in time and, thus, by Ostrogradsky's theorem \cite{Ostrogradsky}, lead to the appearance of ghosts. 

Given these two contrasting examples, a crucial question is then: which kinds of complex scalar field generalizations of the Galileon Lagrangian does supersymmetry allow? We now turn to this question.


\section{Supersymmetric Cubic Galileons} \label{SupersymmetricGalileons}

In this section, 
we will construct all possible supersymmetric Lagrangians involving the product of three fields and four space-time derivatives, in order to see if there might exist inequivalent supersymmetric extensions of the $L_3$ Lagrangian (\ref{L3}). 
We will work in $N=1$ superspace (for a detailed exposition see \cite{Wess:1992cp}). Here, in addition to ordinary four-dimensional bosonic spacetime one adds four fermionic, Grassmann-valued dimensions. These have coordinates $\th_\a$ and $\bar\th_{\dot\a}$, transforming as a two-component Weyl spinor and conjugate Weyl spinor respectively.  One can then define the superspace derivatives
\be
D_\a = \frac{\pt}{\pt\th^\a} + \I \sigma_{\a \dot\a}^\mu \bar\th^{\dot\a} \pt_\mu, \qquad \bar{D}_{\dot\a} = - \frac{\pt}{\pt\bar\th^{\dot\a}} - \I \th^\a \sigma_{\a\dot\a}^\mu \pt_\mu
\label{new1}
\ee
which satisfy the supersymmetry algebra
\be 
\{ D_\a, \bar{D}_{\dot\a} \} = -2\I \sigma^\mu_{\a\dot\a} \pt_\mu \ .
\label{mtblanc1}
\ee
Any superfield can be expanded in the anti-commuting coordinates $\th,\bar\th,$ with the expansion terminating at order $\th\th\bar\th\bar\th$ because of the Grassmann nature of the fermionic coordinates.  A chiral superfield $\Phi$ is defined by the constraint
 \begin{equation}
\bar{D} \Phi =0 \ .
 \label{burt1}
 \end{equation}
This has the expansion
\begin{eqnarray}
&&\Phi = A(x) + \sqrt{2} \th\chi(x) + \th\th F(x)   \nonumber \\
&&\quad + \I \th\sigma^m \bar\th \pt_m A(x) - \frac{\I}{\sqrt{2}}\th\th\pt_m \chi(x)\sigma^m \bar\th + \frac{1}{4}\th\th\bar\th\bar\th \Box A(x), \label{burt2}
\end{eqnarray}
where $A$ is a complex scalar, $\chi_\a$ is a spin-$\frac12$ fermion and $F$ is a complex auxiliary field. In this paper, we will ignore the fermion. Furthermore, since we are only interested in the structure of kinetic energy terms, we need not introduce a superpotential -- in the absence of  which the $F$ field can, and will, be consistently set to zero. 

What makes superspace so useful is that the top component (that is, the $\th\th\bar\th\bar\th$ component) of a superfield transforms under supersymmetry 
into a total spacetime derivative. Hence, one can use this top component to construct supersymmetric Lagrangians. The top component can be isolated by integrating the superfield Lagrangian over superspace with $d^{4}\theta=\d^2\th \d^2\bar\th$ or, alternatively, by acting on it with $D^2 \bar{D}^2$. The supersymmetry algebra (\ref{mtblanc1}) then implies that the top component of a superfield will contain two additional spacetime derivatives compared to its lowest component or compared to the superfield expression itself. For example, ordinary two-derivative scalar field theories are obtained by isolating the top component of the K\"{a}hler potential, which is an hermitian function of the chiral superfield $\Phi$ and its hermitian conjugate $\Phi^{\dagger}$ involving no spacetime derivatives. 

In our case, we are interested in Lagrangians involving the cubic product of a scalar field and four spacetime derivatives. This means that we should consider all possible superfield expressions involving the cubic product of a chiral superfield and {\it two} spacetime derivatives (and linear combinations of all such terms). The superfield Lagrangians of potential interest are straightforward to write down. They are given by the $\th\th\tb\tb$ components of the following expressions (where derivatives act only on the immediately following superfield):
\bea
&& \p^\mu \Phi \p_\mu \Phi \Phi + h.c. \label{x} \\
&& \pt^\mu \Phi \pt_\mu \Pd \Phi + h.c. \label{y} \\
&& \pt^\mu \Phi \pt_\mu \Phi \Pd + h.c. \label{z}
\eea
All other terms of potential interest can be related to these via linear combinations and using integration by parts. 

One might be concerned that there could be other allowed terms involving the superspace derivatives $D_\a $ and $\bar{D}_{\dot\a} $ in \eqref{new1}. Once again, however, upon integration by parts, using the algebra \eqref{mtblanc1} and the chiral superfield constraint \eqref{burt1}, it follows that these are always equivalent to some linear combination of \eqref{x},\eqref{y} and \eqref{z}. As a concrete example, consider the term
\begin{equation}
\int d^4 x d^4 \theta \bar{D}_{\dot{\alpha}} D^2 \Phi \bar{D}^{\dot\alpha} \Pd \Phi \ .
\label{new2}
\end{equation}
Using integration by parts, algebra \eqref{mtblanc1} and the chiral constraint \eqref{burt1} this becomes
\bea
&& \int d^4 x d^4 \theta \bar{D}_{\dot{\alpha}} D^2 \Phi \bar{D}^{\dot\alpha} \Pd \Phi \\
&=& \int d^4 x d^4 \theta (-\bar{D}^2D^2 \Phi) \Pd \Phi \\
&=& \int d^4 x d^4 \theta (-16 \Box \Phi) \Pd \Phi \\
&=& \int d^4 x d^4 \theta [16 \pt^\mu \Phi \pt_\mu \Pd \Phi + 16 \pt^\mu \Phi \pt_\mu \Phi \Pd]
\eea
and, hence, is simply a linear combination of \eqref{y} and \eqref{z}, as claimed. It is straightforward to show that this is always the case.

Having established this, let us systematically discuss the Lagrangian associated with each of the three supersymmetric terms \eqref{x},\eqref{y} and \eqref{z}. First consider \eqref{x}. Note that this is the only one of the three terms that  can possibly lead to the complex Galileon $L_3^{\mathbb{C}}$ given in \eqref{hello1} of the previous section. This follows from the fact that it is the sole term containing only $\Phi$'s or only $\Phi^\dagger$'s in a single term. Hence, it appears that this might be a suitable supersymmetric extension of the $L_{3}$ Lagrangian with purely second order equations of motion. However, the chirality of $\Phi$ immediately implies that the supersymmetric Lagrangian associated with \eqref{x} is, in fact, zero. To see this, instead of integrating over $\d^4 \th,$ one can make use of the Grassmann nature of the $\th,\tb$ coordinates and replace $\d^4 \th$ by a $D^2 \bar{D}^2$ derivative of the corresponding superfield expression. Since $\bar{D}$ commutes with partial derivatives, it immediately follows that superfield expressions constructed exclusively out of $\Phi$'s and partial derivatives must vanish, since the $\bar{D}$ derivative will necessarily act on a chiral field $\Phi$ thus yielding zero. That is, the supersymmetric action associated with \eqref{x} is
\begin{equation}
\int \d^4 x \d^4 \th \pt^\mu \Phi \pt_\mu \Phi \Phi =0 \ .
\label{new3}
\end{equation}
Note that this argument relies solely on holomorphicity and, thus, also extends to potential supersymmetric extensions of complex Galileons with higher powers of fields, such as $L_4^\mathbb{C}$ and $L_5^\mathbb{C}$.

It follows that we are left  with only two possible supersymmetric extensions of the $L_{3}$ Lagrangian--namely, with integrands \eqref{y} and \eqref{z}. 
These are 
\be
\int \d^4 x \d^4 \th \pt^\mu \Phi \pt_\mu \Pd \Phi = \int \d^4 x \, \big(-A\Box A \Box A^* - \Box A^* (\p A)^2\big) \label{Susy1} 
\ee
and
\be
\int \d^4 x \d^4 \th \pt^\mu \Phi \pt_\mu \Phi \Pd = \int \d^4 x \,  \Box A^* (\p A)^2 \label{Susy2}
\ee
respectively, plus their hermitian conjugates.
Note that we have used integration by parts to simplify these terms as much as possible. Let us first examine the action given in \eqref{Susy1}. We immediately see that this term is {\it not} an appropriate extension of the $L_{3}$ Galileon Lagrangian. This follows from the fact that, when the scalar $\xi$ is set to zero, this Lagrangian does not reduce to $L_{3}$ and in fact results in a fourth-order equation of motion for $\phi$.
Hence, we are left with a single possible supersymmetric extension of the $L_3$ Galileon Lagrangian, namely the real part of (\ref{Susy2}). We note that this Lagrangian is equivalent to the supersymmetric Galileon Lagrangian used in \cite{Khoury:2011da}. Thus, we define the supersymmetric extension of $L_3$ as
\bea
L_3^{SUSY} &\equiv& -\frac{1}{\sqrt{2}} \int \d^4 \, \th \pt^\mu \Phi \pt_\mu \Phi \Pd + h.c. \nn \\ &=& -\frac{1}{\sqrt{2}}  \Box A^* (\p A)^2 + h.c. \nn \\ &=&  -\frac12 \big((\p\phi)^2 \Box \phi - (\p\xi)^2 \Box \phi +2 \p\phi \cdot \p\xi \Box \xi\big) \ . \label{name4}
\eea
Compared to the complex Galileon (\ref{L3C}), only the sign of the last term has changed! Nevertheless, this has profound consequences, since the resulting equations of motion are now of third order in derivatives. They read
\bea
0 &=& (\Box \phi)^2 - \phi^{,\mu\nu} \phi_{,\mu\nu} + (\Box \xi)^2 + \xi^{,\mu\nu} \xi_{,\mu\nu} + 2 \xi_{,\mu} \xi_{,\nu}{}^{\nu\mu},\\
0 &=& \xi^{,\mu\nu} \phi_{,\mu\nu} + \xi_{,\mu} \phi_{,\nu}{}^{\nu\mu}.
\eea
As one can clearly see, it is the presence of the second scalar $\xi$ that induces the dangerous higher-derivative terms. That is, $L_{3}^{SUSY}$ in \eqref{name4}, similarly to the second of our concrete examples given in \eqref{again}, has higher-order equations of motion. We will show explicitly in the next section that the presence of these higher derivatives leads to the appearance of a ghost. 


\section{Hiding From the Ghost} \label{Ghost}

We would now like to explicitly demonstrate the ghost degree of freedom in $L_3^{SUSY}.$ The presence of a ghost is already implied by Ostrogradsky's theorem \cite{Ostrogradsky} and we will, in fact, analyze a supersymmetric version of $L_4$ from this point of view in the following section. Nevertheless, we prefer to also analyze the Lagrangian $L_3^{SUSY}$ directly, both because it is instructive to see the ghost appearing at the level of the Lagrangian and because such an analysis elucidates in what regime the ghost can be harmless. For this purpose, it suffices to look at the time-derivative terms in the Lagrangian, since it is these that are associated with ghosts. Adding a canonical kinetic term $L^{SUSY}_2 = \int \d^4 \th \Phi \Pd = -\p^\mu A \p_\mu A^*$, as well as an overall constant $c_3$ in front of the $L_3^{SUSY}$ Lagrangian, the Lagrangian of interest becomes
\be
L_{2+3}^{SUSY} \equiv  L^{SUSY}_2 + c_3 L^{SUSY}_3= \frac12 \dot\phi^2 + \frac12 \dot\xi^2 + c_3 \dot\xi^2 \ddot{\phi}, 
\ee
where we have integrated by parts in order to place all double derivatives on $\phi$ rather than $\xi$. Note that this is a completely arbitrary choice and does not reduce the generality of our analysis. We consider a time-dependent background and would like to study perturbations around it. Thus, we define
\be
\phi = \bar\phi(t) + \delta \phi(x^\mu), \qquad \xi = \bar\xi(t) + \delta \xi(x^\mu).
\ee
Even though the perturbations depend on both time and space, we will only be interested in the time dependence here. To quadratic order in fluctuations, the Lagrangian then becomes
\be
L^{SUSY}_{2+3~\rm{quad}} = \frac12 (\dot{\delta\phi})^2 + \frac12 (1 + 2 c_3 \ddot{\bar{\phi}})  (\dot{\delta\xi})^2 + 2 c_3 \dot{\bar{\xi}} \, \dot{\delta \xi}\ddot{\delta\phi}.
\ee
By defining a new fluctuation variable
\be
\dot{\delta b} \equiv \dot{\delta \xi} + \frac{2 c_3 \dot{\bar{\xi}}}{1 + 2 c_3 \ddot{\bar{\phi}}} \ddot{\delta \phi} \ ,
\ee
the quadratic Lagrangian can then be diagonalized to become
\be
L^{SUSY}_{\rm{2+3~quad}} = \frac12 (\dot{\delta\phi})^2 + \frac12 (1 + 2 c_3 \ddot{\bar{\phi}})  \big((\dot{\delta b})^2 - \frac{4 c_3^2 \dot{\bar{\xi}}^2}{(1 + 2 c_3 \ddot{\bar{\phi}})^2} (\ddot{\delta\phi})^2\big). \label{Lquaddiag}
\ee
Note that $(\dot{\delta b})^2$ and $(\ddot{\delta\phi})^2$ enter with opposite signs and, hence, one of these two terms is ghost-like\footnote{This ghost was not seen in \cite{Khoury:2011da} because in that paper the perturbation analysis was performed solely around $\bar{\xi} = constant$ backgrounds.}. Assuming that the factor $(1 + 2 c_3 \ddot{\bar{\phi}})$ is positive, the ghost then resides in $\ddot{\delta\phi}$. As the Lagrangian shows, the significance of the ghost is essentially controlled by the size of $c_3 \dot{\bar{\xi}}.$ This can be confirmed by looking at the dispersion relation of $\delta \phi.$ 
If one denotes the four-momentum of $\delta\phi$ by $p_\mu,$ then the associated dispersion relation is given by
\be
p_0^2 \big(1-\frac{4c_3^2\dot{\bar{\xi}}^2}{(1+ 2 c_3 \ddot{\bar{\phi}})}p_0^2\big) = 0 \ ,
\ee
where we have assumed that $\dot{\xi}$ and $\ddot{\phi}$ are slowly varying. The mass $m$ is defined via $p^2 = - p_0^2 = - m^2$ and, hence, the dispersion relation implies that $\delta\phi$ consists of two modes. The first is a massless mode which arises from the ordinary correct-sign kinetic term. The second is the ghost, which has a mass
\be
m_g^2 = \frac{(1+ 2 c_3 \ddot{\bar{\phi}})}{4c_3^2\dot{\bar{\xi}}^2} .
\ee
Note that, as there is an overall wrong sign for the ghost in the Lagrangian, this mass is formally tachyonic. However, it is important to realize that this mass does not arise from a potential, but rather from the kinetic term $(\dot{\delta\phi})^2.$ The implication is that this mass does not indicate a time scale over which the (perturbative) vacuum becomes unstable, but rather an energy scale associated with the ghost. In other words, as long as we are considering fluctuations with energy below $m_g,$ the ghost does not get excited. From an effective field theory point of view, we are protected from the catastrophic instabilities associated with the ghost if we take the cut-off $\Lambda$ of the effective field theory to lie below $m_g.$ At the same time, we must ensure that the background itself, that is, $\dot{\bar{\xi}},$ remains within the range of validity of the effective theory. Hence, an additional requirement is that $|\dot{\bar{\xi}}| < \Lambda^2,$ and similar inequalities must also hold for higher time derivatives of $\xi.$  Together with the requirement $\Lambda < m_g$, this implies that we must impose (assuming $|c_3 \ddot{\bar{\phi}}| \ll 1$)
\be
|\dot{\bar{\xi}}| < \frac{1}{|c_3|^{2/3}}, \quad |\ddot{\bar{\xi}}| < \frac{1}{|c_3|}, \quad \dots
\ee
in order to safely suppress the ghost. Thus, as expected, for general backgrounds one must take the prefactor of the Galileon term to be small for consistency.


\section{$L_4$ and a Hamiltonian Analysis}

In this section, we discuss the four-field Galileon Lagrangian $L_4$ and one of its possible supersymmetric extensions. This Lagrangian was presented in Eq. (\ref{L4}) for a single real scalar $\phi.$ Many inequivalent supersymmetric extensions of this Lagrangian exist. We will study one illustrative example here, which can be obtained after rewriting the Lagrangian using integration by parts:
\bea
L_4 &=& -\frac12 (\p\phi)^2 \big((\Box \phi)^2 - \phi^{,\mu\nu} \phi_{,\mu\nu}\big) \nn\\ &=& -\frac14 \pt^\mu (\p\phi)^2 \p_\mu (\p\phi)^2 + \frac12 \pt^\mu \phi \pt_\mu (\p\phi)^2 \Box \phi \ . \label{bud1}
\eea
Making use of the ``building blocks'' \cite{Khoury:2011da,Khoury:2010gb}
\be
D\Phi D \Phi = -4 \tb\tb (\p A)^2, \qquad D^2 \Phi = -4 \tb\tb \Box A,
\ee
it is straightforward to write a supersymmetric extension of \eqref{bud1} given by
\bea
L_4^{SUSY} &=& \int \! \d^4 \th \Big( \!\! -\! \frac{1}{32} \pt^\mu (D\Phi D\Phi) \p_\mu (\bar{D}\Pd \bar{D} \Pd) +\frac{1}{16} \pt^\mu \Phi \pt_\mu (D\Phi D \Phi) \bar{D}^2 \Pd\Big) + h.c. \\ &=& - \pt^\mu (\p A)^2 \pt_\mu (\p A^*)^2 + \p^\mu A \p_\mu (\p A)^2 \Box A^* + \p^\mu A^* \p_\mu (\p A^*)^2 \Box A \ .
\eea
As in the previous $L_3^{SUSY}$ analysis, we will add this term--now with a coefficient $c_4$--to the standard kinetic term $L^{SUSY}_2 = \int \d^4 \th \Phi \Pd = -\p^\mu A \p_\mu A^*.$ Since we are primarily interested in the issue of ghosts, we need only consider the time-dependent part of the resulting Lagrangian. This is given by
\bea 
L^{SUSY}_{2+4} &\equiv& L^{SUSY}_2 + c_4 L^{SUSY}_4 \\ &=& -\p^\mu A \p_\mu A^* + c_4\big(- \pt^\mu (\p A)^2 \pt_\mu (\p A^*)^2 + \p^\mu A \p_\mu (\p A)^2 \Box A^* + \p^\mu A^* \p_\mu (\p A^*)^2 \Box A\big) \nn \\
&=& \frac12 \dot\phi^2 + \frac12 \dot\xi^2 + 2 c_4 \dot\xi^2 \ddot\xi^2 + 2 c_4 \dot\xi^2 \ddot\phi^2.
\eea
Again, the higher-derivative nature of the Lagrangian is manifest. The {\it fourth-order} equations of motion that follow from this Lagrangian are
\bea
0 &=& -\ddot\phi + 4 c_4 \frac{\d^2}{\d t^2}(\dot\xi^2 \ddot\phi), \label{EL1}\\ 0 &=& -\ddot\xi + 4 c_4 \frac{\d^2}{\d t^2}(\dot\xi^2 \ddot\xi) -4 c_4 \frac{\d}{\d t}(\dot\xi \ddot\xi^2 + \dot\xi \ddot\phi^2).\label{EL2}
\eea
It is instructive to carry out a Hamiltonian analysis of this theory. In our presentation we will follow the detailed treatment of \cite{Woodard:2006nt} -- for a Hamiltonian treatment of Galileons in general see also \cite{Zhou:2010di}. For our canonical coordinates, we will choose $\phi,\xi$ as well as $a\equiv \dot\phi$ and $b\equiv \dot\xi.$ The corresponding momenta are
\bea
\pi_\phi &\equiv& \frac{\p L_{2,4}}{\p \dot\phi} - \frac{\d}{\d t}\frac{\p L_{2,4}}{\p \ddot\phi} = \dot\phi - 4 c_4 \dot\xi^2 \frac{\d^3}{\d t^3}\phi - 8 c_4 \dot\xi \ddot\xi \ddot\phi, \label{piphi}\\ \pi_\xi &\equiv& \frac{\p L_{2,4}}{\p \dot\xi} - \frac{\d}{\d t}\frac{\p L_{2,4}}{\p \ddot\xi} = \dot\xi - 4 c_4 \dot\xi^2 \frac{\d^3}{\d t^3}\xi - 4 c_4 \dot\xi \ddot\xi^2 + 4 c_4 \dot\xi  \ddot\phi^2, \label{pixi}\\ \pi_a &\equiv& \frac{\p L_{2,4}}{\p \ddot\phi} = 4 c_4 \dot\xi^2 \ddot\phi, \\ \pi_b &\equiv& \frac{\p L_{2,4}}{\p \ddot\xi} = 4 c_4 \dot\xi^2 \ddot\xi.
\eea
The Hamiltonian is given by $H = \dot\phi \pi_\phi + \dot\xi \pi_\xi + \dot{a} \pi_a + \dot{b} \pi_b - L^{SUSY}_{2+4},$
which can be re-expressed in terms of the canonical coordinates and momenta as
\be
H = a \pi_\phi + b \pi_\xi + \frac{1}{8c_4} (\frac{\pi_a}{b})^2 + \frac{1}{8c_4}(\frac{\pi_b}{b})^2 - \frac12 a^2 - \frac12 b^2. \label{Hamiltonian}
\ee
Note that this expression is regular at $b=0$ since $\pi_a$ and $\pi_b$ both contain factors of $b^2.$ To check the consistency of this analysis, one should verify that the Hamilton evolution equations $\dot\phi = \frac{\p H}{\p \pi_\phi},~\dot\xi = \frac{\p H}{\p \pi_\xi},\dots$ and $\dot\pi_\phi = - \frac{\p H}{\p \phi},~ \dot\pi_\xi = - \frac{\p H}{\p \xi},\dots$ lead to sensible results. In fact, the evolution equations for the coordinates are easily seen to be satisfied. Those for the $\pi_\phi$ and $\pi_\xi$ momenta result in $\dot\pi_\phi = 0 , \, \dot\pi_\xi =0, $ which are equivalent to the Euler-Lagrange equations of motion (\ref{EL1}) and (\ref{EL2}). The two remaining equations are
\be
\dot\pi_a = a - \pi_\phi, \qquad \dot\pi_b = b -\pi_\xi + \frac{1}{4c_4b^3}(\pi_a^2 + \pi_b^2),
\ee
which are equivalent to the definitions of the momenta $\pi_\phi$ and $\pi_\xi$ given in (\ref{piphi}) and (\ref{pixi}). Thus, we may trust our derivation of the Hamiltonian (\ref{Hamiltonian}).
Crucially, the Hamiltonian depends linearly on both $\pi_\phi$ and $\pi_\xi$ and, therefore, can be made arbitrarily positive or negative by choosing appropriately large $a\pi_\phi,b\pi_\xi$ terms. Thus, the energy is unbounded from both above and below. This is a clear indication that this theory, taken literally, does not admit any vacuum at all and is, thus, unphysical. This explicitly demonstrates the presence of a ghost degree of freedom (here, in fact, there are two ghosts), and leads to conclusions similar to those of the Lagrangian analysis performed in Section \ref{Ghost}. 

The higher-derivative terms that we have discussed arise because of the presence of the second real scalar field $\xi$. If we momentarily fix $\dot\xi = b = 0,$ then the Hamiltonian reduces to the simple form 
\be
H_{\dot\xi=0} = \frac12 a^2,
\ee
which is manifestly positive. Thus, by shutting one's eyes to the presence of the second field, one may--mistakenly--think that these theories admit a stable vacuum. Even though this restriction leads to fallacious conclusions when treating the above theory on a {\it fundamental} level, it nevertheless supports the conclusion that from an {\it effective} field theory point of view perturbations of sufficiently low energy around $\dot\xi=0$ backgrounds can be admissible.



\section{Discussion}

The fact that $N=1$ supersymmetric Galileons containing the product of three chiral fields necessarily admit higher-derivative equations of motion implies that these theories contain ghosts. This means that when supersymmetry is included, cubic Galileons, both of the standard and the conformal variety, lose their special status among higher-derivative scalar theories and should be treated in much the same way as other higher-derivative terms. That is to say, they should be regarded as correction terms in a perturbative, effective field theory framework. By extension, our results are also likely to apply to the relevant parts of Horndeski's most general scalar-tensor theory \cite{Horndeski:1974wa}. We stress that our work has been done in the context of minimal $N=1$ supersymmetry. It would be interesting to carry out a similar analysis for extended supersymmetries.

As discussed in the introduction, the brane construction of Galileon Lagrangians suggested that they could arise as the sole constitutents of membrane worldvolume theories 
in string theory--that is, in a well-defined ultraviolet framework. However, when explicit calculations of higher-order corrections to brane dynamics were carried out--in the non-supersymmetric case of AdS space \cite{Khoury:2012dn} and in the $N=1$ supersymmetric context of heterotic M-theory \cite{Ovrut:2012wn,Lukas:1998yy,Lukas:1998tt,Lukas:1998hk,Lukas:1999kt,Donagi:1999jp}--it was found that, in addition to the Galileon terms, other higher-derivative terms occur. These new terms are not naturally suppressed relative to the Galileons and lead to higher-order equations of motion. This paper shows that, with hindsight, this result is unsurprising--since in a full supersymmetric context the cubic Galileon terms themselves already admit higher-derivative equations. 

A final comment: as already mentioned, a supersymmetric version of $L_4$ leading to second-order equations of motion has recently been discovered by Farakos {\it et al.} in \cite{Farakos:2013fne}. Their construction explicitly shows that there is enough freedom in the supersymmetric extensions of $L_4$ to find a linear combination of terms where all higher-derivative terms cancel out in the equations of motion. It is of importance to realize that many of the interesting applications of Galileon theories crucially depend on having several of the Galileon terms, {\it i.e.} cubic, quartic and/or quintic Galileons, present simultaneously. Thus, in the cases where the cubic Galileon is present also, an open question raised by the present work is then whether or not the attractive properties of the most interesting solutions--such as Vainshtein screening or consistent violations of the null energy condition--can be maintained in a supersymmetric perturbative  context. We leave this question for future work.

\acknowledgements

We would like to thank Justin Khoury and Mark Trodden for sharing their wide-ranging knowledge of higher-derivative theories with us, and Fotis Farakos for sharing results about a new supersymmetric version of the quartic Galileons. M.K. and J.L.L. would like to thank the University of Pennsylvania for hospitality while this work was undertaken. M.K. and J.L.L. gratefully acknowledge the support of the European Research Council via the Starting Grant numbered 256994. B.A.O. is supported in part by the DOE under contract No. DE-AC02-76-ER-03071 and the NSF under grant No. 1001296.

\bibliographystyle{apsrev}
\bibliography{masterGC}

\end{document}